\documentstyle[11pt,newpasp,twoside,epsf]{article}
\def\edcomment#1{\iffalse\marginpar{\raggedright\sl#1\/}\else\relax\fi}
\reversemarginpar

\begin{document}
\title{Solar Cycle Variation of the Solar Internal Rotation:
Helioseismic Inversion and Dynamo Modelling}
 \author{S. Vorontsov}
\affil{Astronomy Unit, Queen Mary, University of London,
  Mile End Road, London E1 4NS, UK; and
  Institute of Physics of the Earth, B.Gruzinskaya 10, Moscow
 123810, Russia}
\author{R. Tavakol, E. Covas}
\affil{Astronomy Unit, Queen Mary, University of London,
  Mile End Road, London E1 4NS, UK}
 \author{D. Moss}
 \affil{Department of Mathematics, The University, Manchester M13 9PL, UK}

\begin{abstract}
 We report our first results on comparing the variations
 of the solar internal rotation with solar activity, as predicted
 by non-linear solar dynamo modelling,
 with helioseismic measurements using the SOHO MDI data.
\end{abstract}

\section{Introduction}
The migrating bands of faster and slower rotation of the solar
 surface were first discovered by Howard and LaBonte (1980).
 By analyzing helioseismic data which are now provided by the
 space (SOHO MDI) and ground-based (GONG) projects,
 Howe et al (2000) and Antia and Basu (2000) have found that
 these ``torsional oscillations'' penetrate quite deep into
 the solar interior, to at least 8 percent of solar radius.

 The mechanism responsible for producing the 11-yr solar
 torsional oscillations is thought to be the non-linear
 interaction between the magnetic field and the solar
 differential rotation. Comparing the spatial and temporal
 structure of the torsional oscillations predicted by the
 theoretical modelling with helioseismic measurements
 would allow the calibration of the theoretical models of the
 solar dynamo, leading finally to better understanding
 of the basic mechanisms of solar magnetic activity.

 In this contribution, we address the predictions of a
 two-dimensional axisymmetric mean-field dynamo model
 in a spherical shell, in which the only nonlinearity
 is the action of the azimuthal component of the Lorentz force
 of the dynamo-generated magnetic field on the solar angular
 velocity (Covas et al 2000). The torsional oscillations
 produced in this model are compared with the results of
 helioseismic inversion of the SOHO MDI data, now available
 over almost half of the 11-yr solar activity cycle.

 Helioseismic measurements are based on analyzing the rotational
 splittings of the solar p-mode frequencies. These have different
 sensitivities to the rotation at different depths and latitudes,
 which can distort the actual rotation profiles when they are
 inverted from the data of finite accuracy.
 We address this problem by using artificial inversions.

\section{Torsional oscillations predicted by dynamo
 modelling}
 We consider the variations in the solar internal
 rotation, which are induced directly by the Lorentz force
 in the numerical simulations based on the axisymmetric
 non-linear mean-field dynamo model described in
 (Covas et al 2000).

 The variations of the angular velocity with time, taken from the
 adopted dynamo model (see Tavakol et al., in preparation, for
 details), are shown
 in Fig.~1. The results are represented in a form suitable
 for the direct comparison with seismic measurements (see
 Fig.~2 below). The variations are measured relative to the
 solar minimum, and plotted after 1, 2, 3 and 4 years of
 increasing solar activity. The torsional oscillations
 penetrate to  bottom of the convection zone, and have their
 largest
 amplitudes near its base. At lower latitudes, an accelerating
 ``zonal flow'' is clearly seen, which propagates towards
 the equator.

 \begin{figure}
 \plotone{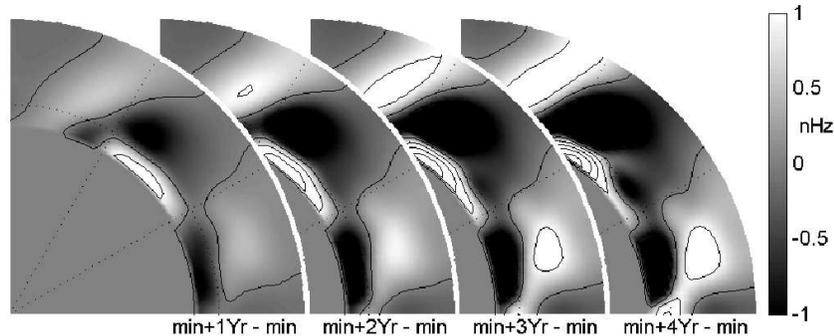}
 \caption{ Variation of the solar rotation
 predicted by the model over one, two, three and four years of
 increasing solar activity. Dotted lines indicate the base of the
 convection zone and the $0^\circ$, $30^\circ$ and $60^\circ$
 latitudes.}
 \end{figure}

 \section{Helioseismic inversion}
 The solar data, represented by the rotational splitting
 coefficients inferred from the SOHO MDI measurements,
 and the inversion technique which we use are described
 in Shou (1999), Vorontsov et al. (2001), Strakhov and
 Vorontsov (2001).

 The results of the inversion are shown in Fig.~2.
 The consecutive 72d data sets (Schou 1999) were averaged
 into one-year data sets to improve the signal-to-noise
 ratio. The inversion has not been applied directly to the
 splitting coefficients, but to their variation relative to
 the first year of measurements, to reduce the effects
 of systematic errors.

 \begin{figure}
 \plotone{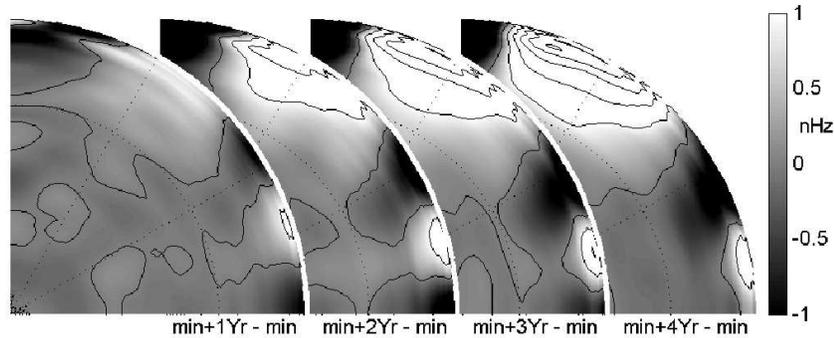}
 \caption{ Variation of the solar internal rotation over one,two,three
 and four years of increasing solar activity, as inferred from the
 SOHO MDI measurements.}
 \end{figure}

 Before comparing these results with model predictions, we
 note a specific feature of the adaptive regularization technique
 (Strakhov and Vorontsov 2001) which was used in the inversion.
 The regularization properties of the inversion are such that
 the response of the solution to the random errors in the input
 data (the noisy component of the solution) is nearly uniform
 over all the approximation domain (the meridional plane).
 If a particular feature in the rotation profile is well below
 the level of
 detectability (determined by data errors), it will not
 be seen in the solution at all; and if its amplitude is close
 to this level, it will appear in the solution with somewhat
 reduced amplitude.

 \section{Artificial inversion}
 To address the question of how the torsional oscillations
 predicted by the model would be ``seen'' in the seismic data,
 an artificial inversion was performed, with using the 2-D
 rotational profile represented by the right-hand panel
 of Fig.~1. The synthetic rotational splitting coefficients
 were calculated for this particular rotation,
 random errors were added, and then the result was
 inverted in exactly the same manner
 as for the real solar data.

 The results are shown in Fig.~3, for inversions with different
 magnitudes of the added noise -- those corresponding to the reported
 observational errors, ten times smaller, and with no errors
 (the error-free inversion indicates the inherent spatial
 resolution of the inversion technique).

 \begin{figure}
 \plotone{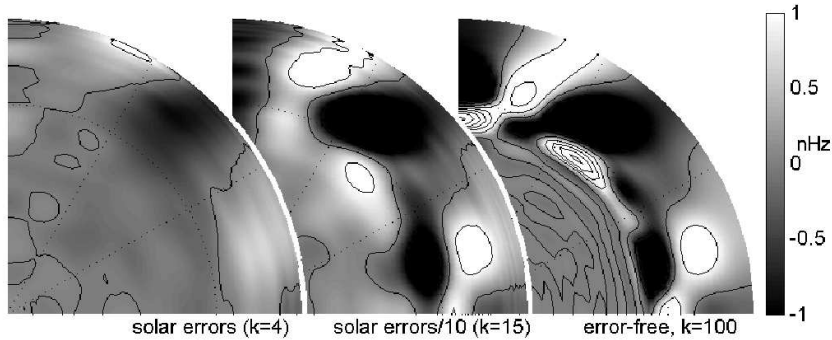}
 \caption{ Artificial inversions of the variation of $\Omega$ predicted
 by the dynamo model over four years of increasing solar activity. From
 left to right: data with noise added corresponding to the reported
 errors of the SOHO MDI one-year data set;
 with noise amplitude reduced by a factor ten; the noise-free inversion.}
 \end{figure}

 The noise-free inversion works well: the only
 noticeable inaccuracy, near the pole, is due to the use of a truncated
 polynomial approximation in $\cos^2(\theta)$. With solar noise
 errors reduced by a factor 10, we are still able to infer
 the variations reasonably well, apart from their detailed
 behaviour near the base of the convection zone.
 In the inversion with solar errors, we can only see the
 variations in the upper half of the convection zone;
 the most interesting features near its base are completely
 buried below the noise level.

\section{Discussion}
 When comparing the torsional oscillations predicted by the
 theoretical
 modelling with helioseismic data, we observe some general features
 which are in common, as well as significant differences.

 Both the model and the observations show that all the convection
 zone, down to its base, is involved in the oscillations.
 Preliminary experiments suggest that this feature remains
 even when density stratification is included.
 The ``zonal flows'' propagate towards the equator at lower
 latitudes and towards the pole at higher latitudes, and the model
 flows show the correct phasing with solar activity.

 The observed low-latitude zonal flow, however, is much more localized
 in depth, compared with model prediction, and closer to the
 surface. The high-latitude accelerating flow is much stronger
 than in the model, has a larger latitudinal extent, and
 is situated at somewhat lower latitudes.

 The differences are hardly surprising; indeed, in this very first
 comparison we made no attempts to tune the model parameters to
 fit the observations, as our interest was in the effect of noise on
 the inversions of a known data set. These differences are  also
 related to
 the inherent uncertainties in, and simplified nature of, the dynamo model:
 for example uniform density is assumed.

 With the current accuracy of the seismic data, we are not able
 to resolve the well-structured variations near the base of the
 convection zone, predicted by the dynamo model.
 We deduce that such structures, if present in the `real' Sun, also
 would not be reliably detected by current techniques.
 We believe that the accuracy of the seismic measurements will
 improve significantly in the near future -- due to improved
 accuracy in the measurement of the rotational splittings in the solar data
 when using the  more sophisticated techniques which are being developed,
  and also from use of the data from ground-based observations,
 together with better coverage of the solar cycle.

 Even with the accuracy which is now available, the seismic
 measurements of the torsional oscillations provide new
 and valuable constraints on the physical modelling of the
 solar dynamo. We believe that we have also shown that dynamo models
 can have an input into the interpretation of the seismic data.

\begin{acknowledgments}
 The Solar Oscillations Investigation (SOI) involving MDI is
 supported by NASA grant NAG5-8878 to Stanford University. SOHO is
 a mission of international cooperation between ESA and NASA. This
 work was supported in part by the UK PPARC under grants
 PPA/G/S/1997/00309 and PPA/G/O/1998/00576.
 DM acknowledges the hospitality of the Astronomy Unit, QM.
\end{acknowledgments}

\end{document}